%% file: main.tex
\documentclass[aps,onecolumn,superscriptaddress,floatfix,nofootinbib,longbibliography]{revtex4-2}

\usepackage{amsmath,amssymb,bm}
\usepackage{graphicx}
\usepackage{booktabs}
\usepackage{array}
\usepackage{xcolor}
\usepackage[margin=1.0in]{geometry}
\usepackage[colorlinks=true,citecolor=black,linkcolor=black,urlcolor=black]{hyperref}

\newcommand{\CrSBr}{CrSBr}
\newcommand{\QSGW}{QS\(G\widehat{W}\)}
\newcommand{\BSE}{BSE}
\newcommand{\dd}{\(d\)-\(d\)}
\newcommand{\ev}{\,\mathrm{eV}}
\newcommand{\rev}[1]{#1}

\makeatletter
\def\frontmatter@abstractheading{\begingroup\centering\large\abstractname\par\endgroup}
\makeatother

\begin{document}

\title{Bright and Dark Excitons in \CrSBr: Local Ligand-Field Character and Band-Coherent Optical Selection Rules}

\author{Swagata Acharya}
\email{swagata.acharya@nlr.gov}
\affiliation{National Laboratory of the Rockies, Golden, CO 80401, USA}
\author{Jessica McDivitt}
\affiliation{University of Colorado Boulder, Boulder, CO 80309, USA}
\author{Dimitar Pashov}
\affiliation{Theory and Simulation of Condensed Matter, King's College London, The Strand, London WC2R 2LS, United Kingdom}
\author{Mark van Schilfgaarde}
\affiliation{National Laboratory of the Rockies, Golden, CO 80401, USA}
\author{Justin C. Johnson}
\affiliation{National Laboratory of the Rockies, Golden, CO 80401, USA}
\author{Jeffrey L. Blackburn}
\affiliation{National Laboratory of the Rockies, Golden, CO 80401, USA}

\date{\today}

\begin{abstract}
Magnetic van der Waals semiconductors such as \CrSBr\ host an intricate exciton landscape. Quasiparticle self-consistent \(GW\) theory with electron-hole ladder vertex corrections (\QSGW) has established the gap and the character of the bright \(X_A\) (\(1.34\,\mathrm{eV}\)) and \(X_B\) (\(1.8\,\mathrm{eV}\)) excitons. Why these bright excitons coexist, within a few tens of meV, with companions orders of magnitude darker has remained unexplained. \rev{Here we show that brightness is fixed by a symmetry of the structure: a mirror-glide exchanges the two Cr atoms of each layer, every exciton has a definite parity under it, and bright and dark partners are its even and odd members. The even sector carries the \(b\)-axis dipole; the odd sector is suppressed by about three orders of magnitude. The dark partners of \(X_A\) and \(X_B\) occupy the same region of \textit{k}-space as the bright states, and dark states such as the one at \(1.50\,\mathrm{eV}\) are dipole-forbidden by inversion symmetry. Frenkel and Wannier-Mott labels describe what an exciton is made of, but brightness is decided by symmetry and by the coherent addition of transition dipoles.} This bare structure is the reference for magnon, phonon and photon couplings.
\end{abstract}

\maketitle

\section*{Introduction}

Magnetic van der Waals (vdW) semiconductors provide an unusual setting for exciton physics because charge, spin, orbital, lattice, and photon degrees of freedom are all active at comparable energy scales \cite{Adak2026Review,Park2026RMPReview}. The discovery of intrinsic 2D ferromagnetism in \(\mathrm{CrI}_3\) and \(\mathrm{Cr}_2\mathrm{Ge}_2\mathrm{Te}_6\) \cite{Huang2017CrI3,Gong2017Cr2Ge2Te6} opened a search for vdW magnets in which optical, magnetic, and transport degrees of freedom can be independently tuned. \CrSBr\ has emerged as one of the most prominent and best-characterised members of this family. First synthesised and characterised decades ago \cite{Beck1990CrSBrStructure,Goser1990CrSBrMagnetism}, it is now established as an A-type antiferromagnetic semiconductor in the orthorhombic \(Pmmn\) structure \rev{with two Cr per layer}. It has ferromagnetic intralayer order, antiferromagnetic interlayer coupling, and a magnetically tunable optical response polarised almost exclusively along the \(b\) axis \cite{Wilson2021Interlayer,Lee2021MagneticOrderSymmetry,Telford2020MagnetoResistance,Telford2022MagnetoTransport,LopezPaz2022HiddenOrder,Klein2022StructureSpin,Yu2023ImagingAFMFM,Ziebel2024AirStable,Cham2022AFMR}.

The two main bright resonances of \CrSBr\ are by now well established. Experimentally they appear as \(X_A\) near \(1.34\,\mathrm{eV}\) and as a broader \(X_B\) manifold near \(1.8\,\mathrm{eV}\) \cite{Wilson2021Interlayer,Klein2023Quasi1D,Smiertka2026FrenkelWannier,Dirnberger2023Nature}. Their wavefunction character, orbital origin, real-space extent, and coupling to the lattice and to the magnon spectrum have been worked out, in close concert with experiment, in a sequence of \QSGW\ studies. Here \(QSGW\) denotes quasiparticle self-consistent \(GW\) \cite{vanSchilfgaarde2006QSGW,Kotani2007QSGW}, in which the one-body Green's function is determined together with the static, Hermitised self-energy so that the result is independent of the DFT starting point; \QSGW\ is its vertex-corrected extension \cite{Cunningham2023QSGWhat}, in which ladder diagrams (electron-hole vertex corrections) are added to the polarisability used to build the screened Coulomb interaction \(W\), and the cycle is iterated to self-consistency. These \QSGW\ studies form part of a broader \QSGW\ programme that has been applied across a range of correlated insulators and two-dimensional vdW magnets \cite{Acharya2021CrX3Bands,Acharya2022CrX3,Acharya2023Colors,Grzeszczyk2023SpinPumping,Belvin2021NiPS3,Adak2026Review}. These works established that (i) the paramagnetic and antiferromagnetic electronic structure of \CrSBr\ \rev{is described by a quasiparticle gap close to \(2\,\mathrm{eV}\), well above the smaller DFT and conventional \(G_0W_0\) values} \cite{Bianchi2023Paramagnetic,Watson2024Exchange,Smolenski2025LargeExciton}; (ii) \(X_A\) and \(X_B\) are strongly bound and have qualitatively different Frenkel-like and Wannier-like wavefunction character, with distinct orbital fingerprints on the Cr \(d\) and ligand-derived bands \cite{Smiertka2026FrenkelWannier}; (iii) an additional family of magnetically confined surface and bulk excitons appears in finite slabs and is accessible to near-field probes \cite{Shao2025SurfaceBulk}; (iv) these excitons couple strongly to photons, entering the strong- and ultrastrong-coupling regime with light and supporting hyperbolic and self-hybridised exciton-polaritons whose dispersion is itself tuned by the underlying magnetic order \cite{Dirnberger2023Nature,Dirnberger2023Hyperbolic,Huang2023Polaritons}; and (v) the same excitons mediate non-trivial couplings to magnons and to the lattice that have now been seen in ultrafast and resonant-driving experiments \cite{Datta2025MagnonMediated,Meineke2024UltrafastExciton,Warshauer2025PopulationInversion,Pawbake2023SpinPhonon}. Taken together, these studies provide a microscopic and experimentally validated picture of \(X_A\) and \(X_B\).

This consensus did not exist five years ago. The first generation of \CrSBr\ exciton papers adopted the language of band-edge Wannier-Mott excitons, labelling the lowest peak as a \(1s\) state of a Rydberg series. Binding energies and effective masses were interpreted from that Rydberg template \cite{Wilson2021Interlayer,Klein2023Quasi1D} that employed an artificially small single-particle gap inherited from DFT and DFT-based \(G_0W_0\) estimates. \QSGW\ corrected this on both sides simultaneously: the paramagnetic and antiferromagnetic gaps are substantially larger than DFT and \(G_0W_0\) values \cite{Bianchi2023Paramagnetic,Watson2024Exchange}, and the exciton binding energies are correspondingly larger \cite{Smiertka2026FrenkelWannier}. The lowest \(X_A\) is therefore not a weakly bound hydrogenic state but a spatially compact object, centred on the Cr and mostly of \dd\ character, and the higher \(X_B\) is a more delocalised excitation, but physically distinct rather than a \(2s\) member of the same ladder. \rev{These gap and binding-energy scales have since been confirmed} by direct ARPES of the relevant valence and conduction manifolds \cite{Bianchi2023Paramagnetic,Watson2024Exchange} and by magneto-optical experiments that resolve \(X_A\) and \(X_B\) as physically separate excitons with distinct symmetry, polarisation, and magnetic-field response \cite{Smiertka2026FrenkelWannier,Dirnberger2023Nature}. They have also been independently reproduced by self-consistent \(GW\) calculations combined with potassium-doping ARPES measurements that place the electronic gap exceeding 1.9\,eV~\cite{Watson2024Exchange} and within \(\sim\!0.05\,\mathrm{eV}\) of \(2\,\mathrm{eV}\) \cite{Smolenski2025LargeExciton}. The current interpretation -- well supported by these mutually consistent theoretical and experimental constraints -- is that strongly bound \(X_A\) is closer to the local ligand-field/Frenkel limit, while more weakly bound \(X_B\) is closer to a Wannier-like intersite, ligand-assisted limit \cite{Smiertka2026FrenkelWannier,Datta2025MagnonMediated,Shao2025SurfaceBulk}, and neither is a member of an effective-mass hydrogenic ladder.

The optical spectrum measured on a real \CrSBr\ sample is never the bare excitonic spectrum. Even at the lowest accessible temperatures, the peaks are dressed by magnon and phonon sidebands \cite{Pawbake2023SpinPhonon,Datta2025MagnonMediated}, and by strong and ultrastrong exciton--photon coupling. \CrSBr\ is now established as a paradigmatic vdW magnetic exciton-polariton platform in which self-hybridised cavity modes, hyperbolic exciton-polaritons, and magnetically tunable polariton branches reshape the apparent excitonic line shape in finite samples and bare flakes alike \cite{Dirnberger2023Nature,Dirnberger2023Hyperbolic,Huang2023Polaritons}, together with light-induced population redistribution on ultrafast and resonantly driven time scales \cite{Meineke2024UltrafastExciton,Warshauer2025PopulationInversion}. Each of these mechanisms can redistribute spectral weight, generate satellite peaks, and modify the apparent oscillator strength of the underlying excitons. The natural reference against which to interpret all of these effects is the spectrum of bare excitons obtained from the Bethe-Salpeter equation (\BSE) -- the ladder summation of electron-hole vertex corrections in the screened Coulomb channel \cite{SalpeterBethe1951,Strinati1988BSE,Onida2002BSE} -- computed on top of the \QSGW\ band structure. \emph{Bare} \BSE\ means only the statically screened Coulomb electron-hole vertex is retained, and no additional boson (magnon, phonon, or photon) dressing is included on top. \rev{Frameworks that add the dressing inside a \(GW\)-like theory exist for some of these bosons, and they are not the calculations that have been reported for \CrSBr. Phonon dressing of \BSE\ excitons is formulated in \cite{Antonius2022ExcitonPhonon,Paleari2022ExcitonPhonon,Dai2024ExcitonicPolaron}. Exciton-polaritons are obtained from a quantum-electrodynamical extension of the \BSE\ \cite{Latini2019QEDBSE}. Magnons and excitons have been obtained together from one Bethe-Salpeter spectrum in monolayer \(\mathrm{CrI}_3\) \cite{Olsen2021CrI3BSE}, using local-density bands with a scissors shift to a \(GW\) gap; that calculation does not dress the optical exciton with magnons. Model Hamiltonians have been used for magnon-mediated coupling in antiferromagnets \cite{Johansen2019MagnonExciton}. In \CrSBr\ the phonon, magnon, and photon couplings are established experimentally \cite{Lin2024ExcitonPhononCrSBr,Datta2025MagnonMediated,Dirnberger2023Nature}, and the theory used to interpret them is in most cases a model analysis rather than that field-theoretical dressing evaluated on the material. The present work supplies the bare excitons on which these dressings should be built.} To address this knowledge gap, this study analyses the bare exciton spectrum for CrSBr, decomposing its individual eigenstates, and in particular their oscillator strengths and bright/dark partitioning, to provide the fundamental baseline response onto which subsequent magnon, phonon, and photon dressings can be layered.

Our analysis reveals insights that so far have not been addressed - that the well-known \(X_A\) and \(X_B\) bright peaks are accompanied within a few tens of meV by near-degenerate companion states that are several orders of magnitude darker, even though they draw from essentially the same single-particle transition manifold. While the standard interpretive language used in the literature \cite{Wannier1937Excitation,Mott1938PolarCrystals,Frenkel1931LightHeat,Tanabe1954AbsorptionI,Griffith1964TransitionMetalIons,ZhangRice1988CuOxides,Davydov1971Excitons,Kasha1965Aggregates} - Frenkel versus Wannier-Mott, ligand-field \dd, charge-transfer, hydrogenic - tells us what microscopic ingredients make up a given exciton, it says nothing about why two excitons with closely matched microscopic ingredients can differ by many orders of magnitude in oscillator strength.

The dark companion excitons have been observed experimentally as well. Resonant inelastic x-ray scattering (RIXS), whose transitions are not masked by the optical matrix element, has recently resolved a dispersive dark excitonic excitation in \CrSBr\ near \(1.5\,\mathrm{eV}\) \cite{Sears2025RIXSDarkExciton}, \rev{in an energy window where our \QSGW\ spectrum carries several states with vanishing \(b\)-axis oscillator strength. We compare energies only: the measured feature is dispersive, whereas our bare excitons are vertical (\(\mathbf{Q}=0\)) and carry no RIXS cross section, so we describe the agreement as consistent with, not as an identification of, the same parent state.} \rev{Coherent magnon- and phonon-driven transient reflectivity shows a discrete resonance near \(1.46\,\mathrm{eV}\) that is absent from steady-state spectra \cite{Borka2026OpticalInterface}.} Here we show how elementary symmetry arguments explain the stark contrast between nominally similar excitons.

\rev{The Frenkel and Wannier pictures depict opposite limits of electron-hole pair binding. In the Frenkel limit, the reference is the free atom. Atomic multiplets are perturbed by weak coupling to ligand neighbors so excitonic features are essentially untethered from the host band structure. In the Wannier limit, the reference consists of freely propagating, noninteracting electron and hole Bloch states at the conduction band minimum and valence band maximum. Electron and hole interact by analogy with the hydrogen atom, \textit{mutatis mutandis}, with effective masses replacing electron masses and the bare Coulomb interaction screened. The excitonic energy levels are the renormalized Ry series, with binding energy relative to the conduction band edge. An exciton in the Wannier limit is derived from states near the two band extrema, and thus confined to a small region of \textit{k}. The Wannier picture is the traditional way to describe excitons in conventional semiconductors and has been used in many nonmagnetic two-dimensional materials as well, although even nonlocal screening, anisotropic mass and reduced dimensionality modify the ideal hydrogenic series. First-principles \(GW\)-\BSE\ calculations on monolayer transition-metal dichalcogenides have demonstrated a rich diversity of bound exciton states sitting on top of a non-hydrogenic, dimensionally reduced screening profile \cite{Qiu2013MoS2,Qiu2016MoS2Screening}, and the resulting non-hydrogenic Rydberg series was confirmed experimentally for monolayer \(\mathrm{WS}_2\) \cite{Chernikov2014Nonhydrogenic}.}

\begin{figure*}[!htbp]
\centering
\includegraphics[width=\textwidth]{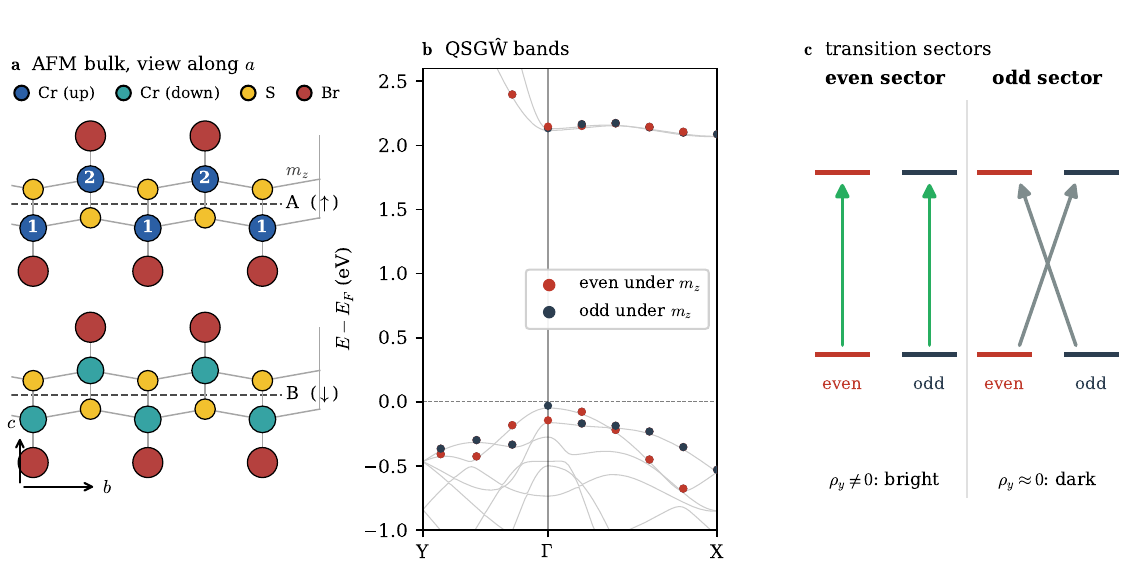}
\caption{\textbf{Structure and symmetry of bulk \CrSBr.} \rev{a) Antiferromagnetic bulk cell viewed along the \(a\) axis: two layers, A (spin up) and B (spin down), each with two Cr atoms (labelled 1 and 2) and S and Br atoms. Blue circles are Cr atoms of layer A, teal circles Cr atoms of layer B, yellow circles S and dark red circles Br. The mirror-glide \(m_z\) (\(z\to -z\) followed by the translation \((\tfrac12,\tfrac12,\tfrac12)\) of the cell) maps each layer onto itself and exchanges Cr~1 and Cr~2. b) \QSGW\ bands along \(\Gamma\)-Y and \(\Gamma\)-X (\(k_z=0\)), with the energy \(E\) measured from the Fermi energy \(E_F\). Mesh points of the two highest valence and two lowest conduction bands are marked by red circles (even under \(m_z\)) and dark blue circles (odd under \(m_z\)); grey lines are the bands. Which band of a pair is higher in energy changes along the path. The quasiparticle gap is about \(2.1\ev\). c) Schematic of the transitions between the two valence levels and the two conduction levels at one \(k\)-point; red bars are even and dark blue bars odd levels. In the even sector (left), green arrows join levels of equal parity and the \(b\)-axis dipole \(\rho_y\) is allowed (bright). In the odd sector (right), grey arrows join levels of opposite parity and the dipole is suppressed by about three orders of magnitude (dark).}}
\label{fig:schematic}
\end{figure*}

We resolve the per-channel optical amplitude of each \QSGW\ exciton in CrSBr directly from its \BSE\ eigenfunction and the underlying single-particle transition-density vector. The brightness is a function of the coherence with which its constituent Bloch transitions add. \rev{The lowest excitons of \CrSBr\ are described to good approximation by a minimal block of the two topmost valence and two lowest conduction branches (Fig.~\ref{fig:schematic}), within the larger set of bands included in the calculation. Bright and dark partners sample similar regions of the Brillouin zone, but they have opposite parity under the mirror-glide operation \(m_z\) that exchanges the two Cr atoms of each layer (Fig.~\ref{fig:schematic}). The even sector carries the \(b\)-axis dipole matrix element; in the odd sector it is smaller by about three orders of magnitude. Bright excitons are therefore even and their dark partners odd (Fig.~\ref{fig:summary}). A second group of dark states is even under \(m_z\) and even (gerade) under inversion, which forbids all three dipole components because the dipole is odd under inversion. These statements are made precise by classifying every exciton by the irreducible representations of the point group of the structure. The classification does not depend on onsite \dd\ weight or on the distribution in \textit{k}-space or real space, and it applies to both Frenkel-like and Wannier-like excitons.}

\section*{Results}

\rev{The calculation is for the bulk crystal in its antiferromagnetic ground state. The magnetic unit cell contains two layers and four Cr atoms (Fig.~\ref{fig:schematic}a), and \(x\), \(y\), \(z\) denote the crystallographic \(a\), \(b\), \(c\) axes.}

\subsection*{Excitons between the Frenkel and Wannier limits}

\CrSBr\ violates several assumptions behind the Wannier-Mott picture. The relevant bands are not weakly correlated \(sp\)-like bands but Cr \(d\)-dominated, ligand-hybridized magnetic bands\rev{: the lowest conduction bands are mainly Cr \(d_{x^2-y^2}\) and \(d_{z^2}\) in character, and the top valence bands mix Cr \(d_{xy}\), \(d_{yz}\) and \(d_{z^2}\) with S and Br \(p\) states, with almost no \(d_{xz}\) weight because of the strong in-plane anisotropy \cite{Smiertka2026FrenkelWannier}. In the excitons \(X_A\) and \(X_B\) the hole is mainly Cr \(d_{yz}\) (with a large S and Br \(p\) share for \(X_B\)) and the electron mainly \(d_{x^2-y^2}\) and \(d_{z^2}\) (Supplementary Fig.~S2 and Table~S2), in line with our earlier orbital analysis \cite{Smiertka2026FrenkelWannier}}. The single-particle gap, when treated at the \QSGW\ level and benchmarked against photoemission, is close to \(2\ev\) and much larger than the smaller gap inferred in early DFT-based and $G_0W_0$ studies \cite{Bianchi2023Paramagnetic,Smiertka2026FrenkelWannier}. As noted earlier, this larger electronic band gap is now corroborated by an independent self-consistent \(GW\) implementation combined with low-temperature K-dosing ARPES \cite{Smolenski2025LargeExciton}, so the choice of starting point and the level of self-consistency are no longer the limiting source of uncertainty. As a consequence of the larger gap, the \(X_A\) and \(X_B\) excitons are both strongly bound, with binding energies of about \(0.8\ev\) and \(0.3\ev\) relative to the quasiparticle gap (Methods), close to the values of roughly \(0.7\ev\) and \(0.3\ev\) below the conduction-band minimum in our earlier work \cite{Smiertka2026FrenkelWannier}. These binding energies are too large, and the orbital content too transition-metal-centered, for the lowest manifold to be viewed as an ordinary Rydberg ladder of weakly bound band-edge states. A realistic picture cannot simply adjust numerical parameters in the effective-mass model, but changes the microscopic identity of the excitons.

The ligand-field/Frenkel language captures the other side of the truth. The \(X_A\) manifold has strong onsite Cr \dd\ character, distributed broadly in \textit{k}-space and compactly in real-space, consistent with the Frenkel limit. But an ideal atomic multiplet or molecular ligand-field picture would miss three essential facts. First, the excitons carry large, highly anisotropic oscillator strengths. Second, the bright and dark states are formed from Bloch-band transitions, not from isolated ionic levels. Third, nearby states with similar local orbital content can have oscillator strengths differing by many orders of magnitude. Thus local orbital character identifies what orbitals compose the exciton (the Frenkel limit applies more to the \(X_A\) exciton than the \(X_B\) exciton) but not whether it is optically bright.

A better analogy for CrSBr, that transcends both the Wannier-Mott and Frenkel pictures, is a correlated magnetic-band-insulator version of a composite local excitation. The Cr-centered \dd\ component, intersite Cr-Cr component, and ligand-assisted \(p\)-\(d\) component are not separate species of excitons, but are components of the same BSE eigenvector \cite{Smiertka2026FrenkelWannier,Datta2025MagnonMediated,Shao2025SurfaceBulk}. Thus the connection to Zhang-Rice physics (which would entail only a single Cr ion and its ligands) is conceptual rather than literal. A nominally local transition-metal excitation becomes a symmetry-adapted, ligand-hybridized object whose low-energy properties cannot be read off from an isolated atomic configuration. For CrSBr, this composite nature is amplified by the orthorhombic lattice, strong \(b\)-axis optical anisotropy, magnetic stacking, and near-degenerate bands.

\subsection*{Optical amplitude and the brightness rule}

\rev{The excitons are eigenstates of the Bethe-Salpeter Hamiltonian in the Tamm-Dancoff approximation, built from the \QSGW\ quasiparticle energies and the statically screened interaction,}
{\begin{equation}
 \sum_{j}\Big[(E_{ck}-E_{vk})\,\delta_{ij}+K_{ij}\Big]A^{S}_{j}=\Omega_S\,A^{S}_{i},\qquad i\equiv(k,v,c),
 \label{eq:bse}
\end{equation}}
\rev{with kernel \(K=K^{\mathrm{x}}-K^{\mathrm{d}}\) (bare exchange minus screened direct electron-hole attraction) and normalised eigenvectors, \(\sum_i|A^S_i|^2=1\). \(\Omega_S\) is the excitation energy of exciton \(S\), a vertical (\(\mathbf{Q}=0\)) optical energy and not a binding energy; the binding energy is the difference between the smallest vertical quasiparticle gap and \(\Omega_S\).}

Let
\begin{equation}
 i \equiv (k,v,c)
\end{equation}
label a vertical transition from valence band \(v\) to conduction band \(c\) at point \(k\) in the Brillouin zone, and let \(A_i^S\) be the BSE eigenvector of exciton \(S\). For polarization direction \(\alpha=x,y,z\), the optical transition-density vector can be written as
\begin{equation}
 \rho_{kvc}^{\alpha}
 =
 \frac{2\,p^{\alpha}_{kcv}}
 {E_{ck}-E_{vk}} ,
 \label{eq:rho}
\end{equation}
where \(E_{ck}\) and \(E_{vk}\) are the quasiparticle eigenvalues and \(p^{\alpha}_{kcv}=\langle c\mathbf{k}|\hat{p}_{\alpha}|v\mathbf{k}\rangle\) is the inter-band momentum matrix element along polarisation \(\alpha\) \rev{(the index \(c\) comes first because the element is \(\langle c|\hat p|v\rangle\); the transition \(v\to c\) carries the label \((k,v,c)\)).} \rev{The denominator is the vertical energy of the quasiparticle bands. In the active window it is never smaller than about \(2.1\ev\). Degeneracies occur only among valence bands or among conduction bands; they leave the denominator unchanged but make the label of each band a choice of basis, which we fix below by rotating each degenerate group into eigenstates of \(m_z\).} The scalar factor \(2p^{\alpha}_{kcv}/(E_{ck}-E_{vk})\) is the momentum matrix element divided by the vertical transition energy. With the energies in Rydberg units the factor 2 converts them to Hartree units, so \(\rho^{\alpha}_{kvc}\) is the inter-band dipole matrix element \(\langle c\mathbf{k}|\hat r_\alpha|v\mathbf{k}\rangle\) in atomic units, a length rather than a momentum; throughout the paper only \(\rho\) appears in the sums over transition channels, the raw momentum matrix element \(p^{\alpha}_{kcv}\) is not summed separately. The optical amplitude is then
\begin{equation}
 D_S^\alpha
 =
 \sum_{kvc}
 A_{kvc}^S
 \rho_{kvc}^{\alpha},
 \label{eq:amp}
\end{equation}
and the oscillator strength is
\begin{equation}
 f_S^\alpha
 =
 \mathcal{N}_{\alpha}
 \left|D_S^\alpha\right|^2 ,
 \label{eq:osc}
\end{equation}
\rev{where \(\mathcal{N}_\alpha=1\). In these units \(f_S^\alpha=|D_S^\alpha|^2\) is the oscillator strength written by the \BSE\ code. The comparisons in this paper are ratios of these strengths, and a constant multiplying every \(f\) cancels. The interference ratio \(\mathcal{I}\) introduced below is built from \(A\) and \(\rho\), before \(\mathcal{N}_\alpha\) enters.} This is the standard \BSE\ structure of an excitonic optical response, rooted in the Hedin equations \cite{Hedin1965GW}, the Bethe-Salpeter equation \cite{SalpeterBethe1951,Strinati1988BSE}, and their first-principles implementations for semiconductors and insulators \cite{HybertsenLouie1986,RohlfingLouie2000,Onida2002BSE,Kotani2007QSGW}.

Equations \eqref{eq:amp} and \eqref{eq:osc} immediately show why a weight-only interpretation can fail:
\begin{equation}
 f_S^\alpha
 =
 \mathcal{N}_{\alpha}
 \sum_{i,j}
 A_i^S \rho_i^\alpha
 \left(A_j^S\rho_j^\alpha\right)^\ast .
 \label{eq:interference}
\end{equation}
All transition channels interfere with all other channels. A dark exciton is not simply a state with no allowed local component; it is a state whose eigenvector is nearly orthogonal to the optical vector,
\begin{equation}
 \sum_i A_i^S \rho_i^\alpha \simeq 0.
\end{equation}
Conversely, a bright exciton may retain sizable locally forbidden character if its allowed intersite, ligand-assisted, or band-hybridized components add coherently.

For physical interpretation one can group transitions into subsets \(g\), for example by some volume in \textit{k}, band pair, or orbital/site character:
\begin{equation}
 C_g^\alpha(S)
 =
 \sum_{i\in g}
 A_i^S \rho_i^\alpha .
\end{equation}
The oscillator strength is then
\begin{equation}
 f_S^\alpha
 =
 \mathcal{N}_{\alpha}
 \sum_{g,h}
 C_g^\alpha(S)C_h^\alpha(S)^\ast .
\end{equation}
The off-diagonal group terms are the mathematical expression of constructive and destructive interference \rev{(for the per-\textit{k} groups \(g=k\) see Fig.~\ref{fig:summary}d)}.

\subsection*{An interference figure of merit for brightness}

The key observable derived from Eqs.~\eqref{eq:amp}-\eqref{eq:interference} is the ratio of the coherent and incoherent sums of the per-channel contributions,
\begin{equation}
 \mathcal{I}_S^\alpha
 =
 \frac{\left|\sum_{kvc} A_{kvc}^S\rho_{kvc}^\alpha\right|^2}
 {\sum_{kvc}\left|A_{kvc}^S\rho_{kvc}^\alpha\right|^2}.
 \label{eq:cohinc}
\end{equation}
\rev{\(\mathcal{I}_S^\alpha\) is the ratio of the coherent sum to an incoherent reference sum. It is not a bound (\(\mathcal I>1\) occurs) and not a brightness threshold: \(f=\mathcal I\sum|A\rho|^2\), so brightness also depends on the size of the reference sum. For example \(X_B\) has \(\mathcal I\approx 10\) and \(f_y=134\), whereas the \(z\)-polarised state at \(1.86\ev\) has \(\mathcal I\approx 5\) and \(f_z=0.19\). \(\mathcal{I}\) also depends on the band basis wherever bands are degenerate, because rotations among degenerate bands change the individual terms but not their sum. The per-\textit{k} variant \(\mathcal I_k\), built from \(C_k=\sum_{vc}A_{kvc}\rho_{kvc}\), is unaffected. We report \(\mathcal I\) in the original band basis and in a symmetry-adapted basis, in which the bands of each degenerate group are rotated into eigenstates of the mirror-glide \(m_z\) introduced below, so that the labels of degenerate bands are fixed by symmetry and not by the numerical diagonalisation.} We evaluate \(\mathcal{I}_S^\alpha\) by reconstructing the optical amplitude of Eq.~\eqref{eq:amp} directly from the \BSE\ eigenvectors and the independent-particle transition-density vector of Eq.~\eqref{eq:rho}. This reconstruction reproduces the oscillator strength reported by the \BSE\ solver for every exciton, in all three polarisation channels and both spin channels, to numerical precision (Methods). Throughout, the dominant optical response is along the crystallographic \(b\) axis (\(\alpha=y\)), so \(\mathcal{I}_S\) without a polarisation index refers to this dominant b-axis channel of state \(S\).

\subsection*{\rev{Symmetry of the structure and parity of the bands}}

\rev{With the Cr spin labels kept, the antiferromagnetic bulk cell has eight spin-preserving symmetry operations: the identity, the three two-fold rotations \(C_{2x},C_{2y},C_{2z}\), inversion \(i\), and three mirrors (those with a translation carry \((\tfrac12,\tfrac12,\tfrac12)\) of the cell). They form a group isomorphic to \(D_{2h}\). Four of them (\(i\), \(C_{2x}\), \(C_{2y}\), \(m_z\)) exchange the two Cr atoms of each layer. The mirror-glide \(m_z\) is the simplest: it maps each layer onto itself, it is a symmetry of an isolated layer, and it maps every \(k\)-point of the calculation onto itself. Because the two layers carry opposite spin and are coupled only by van der Waals interaction, the excitons are essentially intralayer (Supplementary Fig.~S4), and the classification below is that of a single layer.}

\rev{We construct \(m_z\) on the Bloch basis of the calculation (Methods). Rotating the bands within degenerate clusters into eigenstates of \(m_z\) (the symmetry-adapted basis) gives every band a definite parity (Fig.~\ref{fig:schematic}b). The two highest valence bands have opposite parity at every \(k\)-point, so the four transitions between them and the two lowest conduction bands split into exactly two even and two odd transitions at every \(k\), whichever band lies higher in energy. This is why transitions are labelled by parity and not by band index.}

\begin{figure*}[!htbp]
\centering
\includegraphics[width=\textwidth]{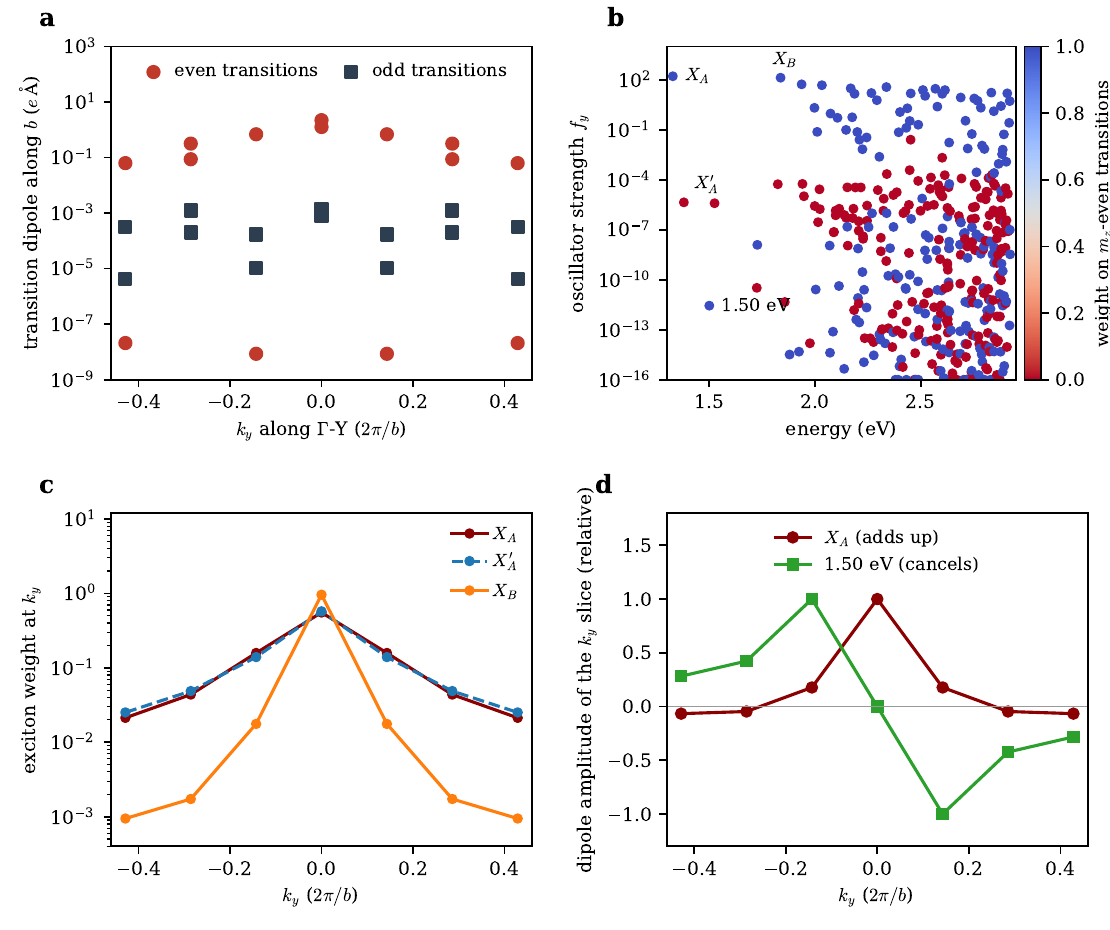}
\caption{\textbf{Brightness and symmetry of the excitons.} \rev{a) Transition dipole along \(b\), in units of \(e\)\,\AA, between the two highest valence and the two lowest conduction bands along \(\Gamma\)-Y, as a function of \(k_y\) (in units of \(2\pi/b\)); red circles are even and dark blue squares odd transitions. At some \(k\)-points one of the two even transitions is weak. b) Oscillator strength along \(b\), \(f_y\), of the lowest two-particle eigenvalues against energy. The colour of each point gives its weight on \(m_z\)-even transitions, from 0 (red) to 1 (blue); the few eigenvalues that could not be classified cleanly are omitted. The bright \(X_A\) and \(X_B\), the partner \(X_A'\) and the state at \(1.50\ev\) are labelled. c) Distribution of the bright \(X_A\) (dark red line with circles), its partner \(X_A'\) (blue dashed line with circles) and \(X_B\) (orange line with circles) over \(k_y\), summed over \(k_x\) and \(k_z\), on a logarithmic scale: \(X_A\) and \(X_A'\) are distributed alike over the zone, whereas \(X_B\) is concentrated at \(\Gamma\). d) Dipole amplitude of the \(k_y\) slices, relative to the largest, for the bright \(X_A\) (dark red line with circles) and for the even, inversion-forbidden state at \(1.50\ev\) (green line with squares). The arbitrary overall phase of each exciton is fixed so that its largest amplitude is real. The sign is physical: slices of the same sign add to the total dipole of the exciton and slices of opposite sign reduce it. The bright \(X_A\) is dominated by the slices around \(\Gamma\) with a small opposite contribution from the outer slices, whereas for the state at \(1.50\ev\) the slices at \(k_y\) and \(-k_y\) cancel pairwise, as inversion requires.}}
\label{fig:summary}
\end{figure*}

\rev{The dipole along \(b\) (and \(a\)) is even under \(m_z\), along \(c\) odd. In the symmetry-adapted basis more than \(99\%\) of the transition-dipole strength along \(b\) (the squared dipole matrix element \(|\rho^y|^2\) summed over all transitions) lies on even transitions, and likewise for \(a\), whereas only \(0.1\%\) of the strength along \(c\) does. For the four transitions of the leading block the odd-sector dipole is much smaller than the even one (Fig.~\ref{fig:summary}a, Supplementary Fig.~S5).}

\subsection*{\rev{Symmetry classification of the excitons}}

\rev{The \BSE\ Hamiltonian commutes with every operation \(\hat g\) of the group, so each non-degenerate exciton at \(\mathbf Q=0\) is an eigenstate of \(\hat g\) with eigenvalue \(\pm1\). We evaluate the characters from the eigenvectors and the band overlaps,
\begin{equation}
 \chi_S(g)=\sum_{k}\sum_{vw,\,cd}\big(A^S_{gk,\,vd}\big)^{*}\,\big(O^{v}_{vw}(k)\big)^{*}\,O^{c}_{dc}(k)\,A^S_{k,\,wc},
 \label{eq:chi}
\end{equation}
where \(O_{ab}(k)=\langle\psi_a(gk)|\hat g|\psi_b(k)\rangle\), for \(g=C_{2z},C_{2y},C_{2x},i\). The dipole components \(x,y,z\) belong to \(B_{3u},B_{2u},B_{1u}\). In the symmetry-adapted basis the weight of an exciton on even transitions is
\begin{equation}
 W_{\mathrm{even}}=\sum_{k}\sum_{(g,h)\,\mathrm{even}}\big|A'_{k,gh}\big|^{2},\qquad A'=U_v^{T}AU_c^{*},\quad \rho'=U_v^{\dagger}\rho\,U_c,
 \label{eq:weven}
\end{equation}
with \(U\) the unitary matrices that diagonalise \(m_z\) within each group of degenerate bands. Because the rotation is unitary, the amplitude \(\sum A'\rho'=\sum A\rho\), and hence the oscillator strength, is unchanged; only its partition into band-resolved terms depends on the basis.}

\rev{Each of the 11 excitons of Table~\ref{tab:states} has characters \(\chi=\pm1\) under every operation and therefore belongs to a single irreducible representation. The bright excitons \(X_A\), \(X_B\) and \(X_C\) (1.33, 1.84, 1.94 eV) belong to \(B_{2u}\), the representation of the \(b\)-axis dipole; the weakly \(x\)-polarised state at 1.73 eV belongs to \(B_{3u}\) and the \(z\)-polarised state at 1.86 eV to \(B_{1u}\). The partners of \(X_A\) (1.38 eV) and \(X_B\) (1.82 eV) are odd under \(m_z\) (\(B_{3g}\)). The states at 1.50 and 1.88 eV are even under \(m_z\) but gerade, so inversion forbids all three dipole components, and the 1.53 eV state and the \(B_{2g}\) state near 1.73 eV belong to representations with no dipole component.}

\rev{Beyond the 11 selected states, we classified the lowest 1200 eigenvalues of the two-particle Hamiltonian (1.33 to 3.34 eV). Only the 34 that lie below the smallest vertical gap are bound excitons; the rest are continuum electron-hole eigenstates. All eigenvalues below 2.0 eV, and \(97\%\) of the 1200, have a well-defined representation; the rest are clusters cut by the edge of the active window. More than \(99.9\%\) of each oscillator strength \(f_x\), \(f_y\), \(f_z\) is carried by eigenvalues in the representation of the matching dipole component, and every classified eigenvalue with \(f\) below \(10^{-8}\) of the maximum lies in a representation without a dipole component.}

\subsection*{Exciton hierarchy between 1.3 and 1.9~eV}

Table~\ref{tab:states} summarises the most relevant excitons. The dominant optical response is along \(b\) (\(y\)) for all listed states except the \(1.73\ev\) state (weakly \(x\)-polarised) and the \(1.86\ev\) state (predominantly \(z\)-polarised).
\begin{table*}[t]
\footnotesize
\caption{\rev{Selected \CrSBr\ excitons from the \QSGW+\BSE\ (the two spin sectors are degenerate). \(E=\Omega_S\) is the \BSE\ excitation energy, not a binding energy (relative to the quasiparticle gap of about \(2.1\ev\) the binding energy is about \(0.8\ev\) for \(X_A\) and about \(0.3\ev\) for \(X_B\)). ``irrep'' is the irreducible representation of \(D_{2h}\) from the characters of Eq.~\eqref{eq:chi}; the dipole components \(x,y,z\) belong to \(B_{3u},B_{2u},B_{1u}\). \(f\) is the oscillator strength in the dominant polarisation (\(y\) unless noted). \(k\mathrm{PR}\) is the \textit{k}-space participation ratio (Methods). \(W_{\mathrm{even}}\) is the weight on \(m_z\)-even transitions (Eq.~\eqref{eq:weven}). \(\mathcal{I}\) is the ratio of Eq.~\eqref{eq:cohinc} in the original band basis and in the symmetry-adapted basis. Oscillator strengths below \(10^{-8}\) of the largest value are not resolved and are given as upper bounds; for these states \(\mathcal I\) carries no information beyond ``dark''.}}
\label{tab:states}
\begin{ruledtabular}
\begin{tabular}{ccccccc}
\(E\) (eV) & irrep & dominant \(f\) & \(k\mathrm{PR}\) & \(W_{\mathrm{even}}\) & \(\mathcal{I}\) (bands) & \(\mathcal{I}\) (adapted) \\
\hline
1.33 & B2u & $f_y\!\sim\!1.6\times10^{2}$ & 42 & $>\!0.9999$ & $40$ & $40$ \\
1.38 & B3g & $f_y\!\sim\!4.4\times10^{-6}$ & 42 & $<\!10^{-4}$ & $2.7\times10^{-4}$ & $0.012$ \\
1.50 & B1g & $f_y<10^{-6}$ & 69 & $>\!0.9999$ & dark & dark \\
1.53 & Au & $f_y\!\sim\!3.9\times10^{-6}$ & 62 & $<\!10^{-4}$ & $1.3\times10^{-5}$ & $9.7\times10^{-3}$ \\
1.73 & B2g & $f_x<10^{-6}$ & 123 & 0.0027 & dark & dark \\
1.73 & B3u & $f_x\!\sim\!2.8\times10^{-4}$ & 125 & 0.9973 & $5.6\times10^{-4}$ & $5.6\times10^{-4}$ \\
1.82 & B3g & $f_y\!\sim\!5.4\times10^{-5}$ & 9 & $<\!10^{-4}$ & $0.023$ & $0.069$ \\
1.84 & B2u & $f_y\!\sim\!1.3\times10^{2}$ & 10 & $>\!0.9999$ & $10$ & $10$ \\
1.86 & B1u & $f_z\!\sim\!0.19$ & 46 & $<\!10^{-4}$ & $5.3$ & $5.3$ \\
1.88 & Ag & $f_z<10^{-6}$ & 48 & $>\!0.9999$ & dark & dark \\
1.94 & B2u & $f_y\!\sim\!54$ & 4 & $>\!0.9999$ & $7$ & $7$ \\
\end{tabular}
\end{ruledtabular}
\end{table*}

\rev{The table organises the excitons into four comparison sets. First, the \(1.33\) and \(1.38\ev\) states are the low-energy bright and dark pair of the \(X_A\) manifold. They belong to \(B_{2u}\) and \(B_{3g}\), that is, they are even and odd under \(m_z\), and they have almost identical \textit{k}-space distributions. Second, the \(1.82\) and \(1.84\ev\) states form the corresponding pair for \(X_B\) (\(B_{3g}\) and \(B_{2u}\)). Third, the \(1.86\ev\) state (\(B_{1u}\)) is the \(z\)-polarised member of a pair with the \(1.88\ev\) state (\(A_g\)), which has no dipole component. Fourth, the \(X_C\) state near \(1.94\ev\) is more \textit{k}-localised (\(k\mathrm{PR}=4\)) and is a separate higher-energy manifold. The dark states at \(1.50\) and \(1.88\ev\) are of a different kind from the odd partners: they are even under \(m_z\), and their amplitude on each \textit{k}-point is not small but cancels between \(k\) and \(-k\), as required by inversion (Fig.~\ref{fig:summary}d).}

\subsection*{\texorpdfstring{Onsite \dd\ weight does not determine brightness}{Onsite d-d weight does not determine brightness}}

The local orbital decomposition gives an important first clue but not a complete selection rule. The bright \(1.33\ev\) exciton has substantial onsite Cr weight, while the bright \(1.84\ev\) exciton has much less \rev{(the electron and hole on the same Cr atom, as reported by the \BSE\ code; Supplementary Fig.~S2)}. Representative site-class weights are
\begin{align}
 W_{\mathrm{onsite\,Cr}}(1.33\ev) &\simeq 0.34,\\
 W_{\mathrm{onsite\,Cr}}(1.84\ev) &\simeq 0.02.
\end{align}
At the same time, the \(1.84\ev\) state has much larger intersite and ligand-assisted character \rev{(inter-site Cr-Cr weight \(0.55\) against \(0.38\), Cr-ligand weight \(0.38\) against \(0.26\))} \cite{Smiertka2026FrenkelWannier,Datta2025MagnonMediated,Shao2025SurfaceBulk}. A purely local ligand-field expectation would therefore suggest that the \(1.84\ev\) state should be brighter, but our calculation gives the opposite ordering: \(f_y(1.33\ev)>f_y(1.84\ev)\).

This is natural in the amplitude language that decomposes an exciton schematically as
\begin{equation}
 |S\rangle
 =
 a_{dd}|dd\rangle
 +
 \sum_m a_m |m\rangle ,
\end{equation}
where \(|dd\rangle\) denotes onsite \dd-like components and \(|m\rangle\) denotes intersite Cr-Cr, Cr-ligand, ligand-Cr, and other band-hybridized components. If
\begin{equation}
 \langle \rho_y|dd\rangle \approx 0,
\end{equation}
then
\begin{equation}
 D_y^S
 =
 \langle \rho_y|S\rangle
 \approx
 \sum_m a_m \langle \rho_y|m\rangle .
\end{equation}
The onsite weight reduces the fraction of the exciton that can directly couple to light, but the brightness is controlled by the coherent sum over the allowed components. Thus a state with sizable onsite \dd\ weight can be bright if its allowed components are phase aligned with the optical vector.

\begin{figure*}[!htbp]
\centering
\includegraphics[width=\textwidth]{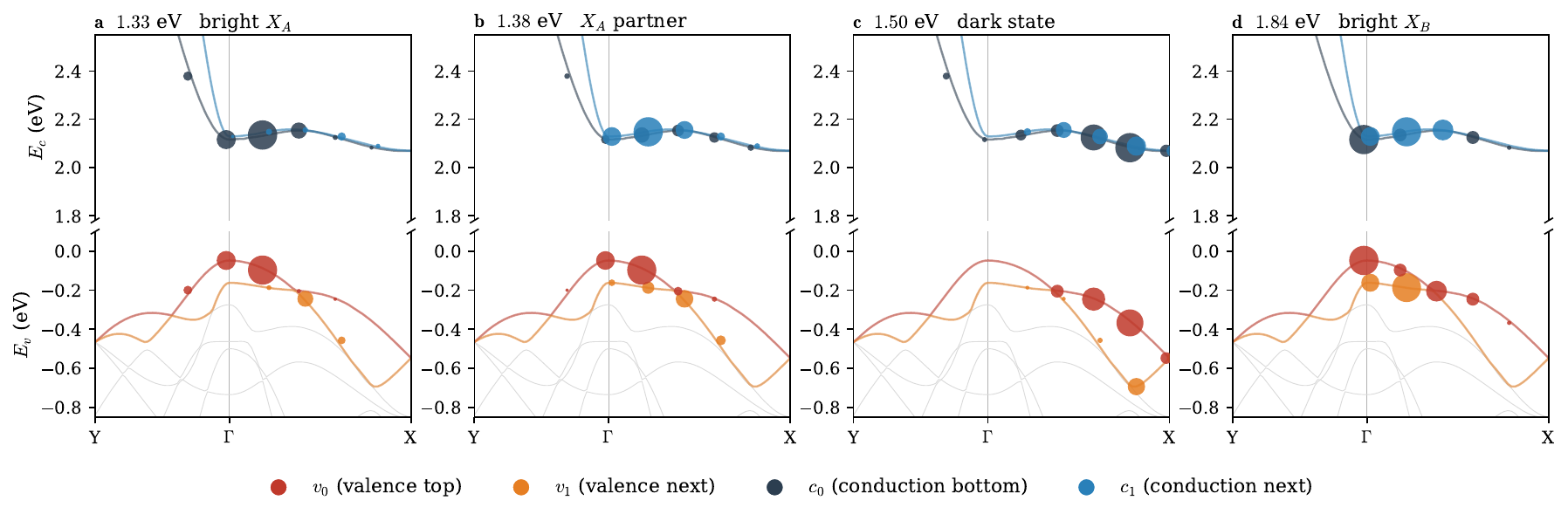}
\caption{\textbf{Band-resolved weight of four excitons along high-symmetry lines.} \rev{BSE weight along \(\Gamma\)-X and Y-\(\Gamma\) (\(k_z=0\) mesh points) for a) the bright \(X_A\) at \(1.33\ev\), b) its partner at \(1.38\ev\), c) the dark state at \(1.50\ev\) and d) the bright \(X_B\) at \(1.84\ev\). Each panel has a broken energy axis: the upper part shows the two lowest conduction bands \(c_0\) (dark blue) and \(c_1\) (light blue) with energy \(E_c\), the lower part the two highest valence bands \(v_0\) (red) and \(v_1\) (orange) with energy \(E_v\), ordered by energy at each \(k\); grey lines are other bands. Marker area scales with the square of the weight relative to the largest weight of the four-band block; weights below \(10\%\) of it are not drawn. Plots of all 11 excitons of Table~\ref{tab:states}, labelled by representation, are in Supplementary Fig.~S1.}}
\label{fig:exbw}
\end{figure*}

This analysis (illustrated in Fig.~\ref{fig:exbw}) explains the \(1.33\ev\) versus \(1.84\ev\) comparison. The \(1.33\ev\) state contains more locally forbidden onsite character, but it is spread over a broader \textit{k}-space manifold (\(k\mathrm{PR}\simeq 40\)). The \(1.84\ev\) state is locally more allowed but much more \textit{k}-localised (\(k\mathrm{PR}\simeq 10\)). \rev{Both are even under the Cr exchange (\(B_{2u}\)) and so are dipole-allowed along \(b\); the \(1.82\ev\) state next to \(X_B\) is odd. Because fewer \textit{k}-space channels participate strongly in \(X_B\), the local allowedness and the total oscillator strength need not be monotonically related.} A useful analogy is molecular exciton theory: an individual monomer transition may be weak or strong, but the aggregate brightness is controlled by the symmetry-adapted sum of transition dipoles. Here the ``monomers'' are not literal molecules; they are Cr-centred, ligand-hybridised band-transition channels distributed over \textit{k} space.

\subsection*{Bright and dark partners occupy similar regions of \textit{k}-space}

\rev{The \(1.33\) and \(1.38\ev\) excitons have almost identical \textit{k}-space distributions: the Bhattacharyya overlap of \(P_S(k)\) and \(P_{S'}(k)\) is \(0.997\), against \(0.84\) for the \(1.33\) and \(1.84\ev\) pair, so the measure discriminates. The \(1.82\) and \(1.84\ev\) pair has overlap \(0.993\). The pairs therefore differ not in where in the zone they live but in their parity. Band-resolved weights (Fig.~\ref{fig:exbw}; Supplementary Fig.~S1) point the same way: the \(X_A\) pair draws its weight from similar bands and a similar region of the zone.}

\rev{In the \(X_A\) pair, the bright state is even and has \(\mathcal I\approx 40\); the \(1.38\ev\) partner is odd, with \(f_y=4\times10^{-6}\). Its distribution over \textit{k}-space is almost that of \(X_A\) (Fig.~\ref{fig:summary}c), but its per-\textit{k} amplitudes \(|C_k|\) are about \(5.5\times10^{3}\) times smaller at every \(k\), and they add coherently across \(k\) (\(\mathcal I_k=16\), against \(22\) for \(X_A\)). The partner is dark because its weight sits on the odd transitions, where the dipole matrix element is small. In the original band basis its \(\mathcal I\) is \(3\times10^{-4}\) and rises to \(1\times10^{-2}\) in the adapted basis, since the individual terms of the sums depend on the choice of basis within degenerate band pairs.}

\rev{The bright \(X_A\) and \(X_B\) excitons both belong to \(B_{2u}\), so symmetry does not distinguish them. They differ in how the electron-hole pair is distributed: the on-site Cr weight is \(0.34\) against \(0.02\), the \textit{k}-space participation ratio is \(42\) against \(10\), and the binding energies are about \(0.8\) and \(0.3\ev\).} The \(1.50\ev\) state, by contrast, has its weight pushed away from \(\Gamma\) along \(\Gamma\)-X, a useful reminder that a direct \(\mathbf{q}{=}0\) exciton does not require its electron and hole to be drawn from \(\Gamma\): momentum conservation only enforces that they be taken from the \emph{same} crystal momentum, anywhere in the Brillouin zone, and the BSE eigenvector is free to concentrate that weight wherever the screened electron-hole interaction makes it energetically favourable \cite{Smiertka2026FrenkelWannier}.

\subsection*{\rev{Parity structure of the four-band block}}

\rev{The calculation includes 26 valence and 9 conduction bands, and the parity classification applies to every transition among them. The lowest excitons are dominated, however, by the two highest valence bands, \(v_\pm\), and the two lowest conduction bands, \(c_\pm\), labelled by their parity under \(m_z\). This block carries \(45\)--\(65\%\) of the weight of \(X_A\) and its partners and about \(95\%\) of \(X_B\) and \(X_C\), but only \(15\)--\(25\%\) of several higher states (Supplementary Table~S1); the rest lies on other bands of the window and follows the same parity rule. The block is therefore a minimal description, not a complete one: it contains the even and odd transitions that decide bright versus dark. The transition vector \(\rho^y_{kvc}\) connects like parities,
\begin{equation}
 \rho^{y}_{v_s c_{s'}}\simeq 0\quad\text{for}\quad s\neq s',
 \label{eq:rhoparity}
\end{equation}
with a measured median ratio of odd to even magnitude of about \(10^{-3}\) (Fig.~\ref{fig:summary}a). In the amplitude \(D_y=\sum A\rho\) of Eq.~\eqref{eq:amp} only the even-sector amplitudes \(A_{++}\) and \(A_{--}\) then contribute,
\begin{equation}
 D_{y}\simeq A_{++}\rho^{y}_{++}+A_{--}\rho^{y}_{--},
 \label{eq:Deven}
\end{equation}
and the odd-sector amplitudes \(A_{+-}\), \(A_{-+}\) are nearly invisible to a \(b\)-polarised field. An exciton in the even sector is bright, one in the odd sector is dark, and because the valence pair is always one even and one odd band, every four-band block contains two transitions of each parity (subsection ``Symmetry of the structure and parity of the bands''), which fixes a bright-and-dark pairing for the lowest excitons. At higher energy the weight is spread over more bands and the characters of Eq.~\eqref{eq:chi} give the representation.}

\rev{The parity assigned by \(m_z\) is a symmetry label of the structure, which exchanges the two Cr atoms of each layer, and not a projection onto an orbital of one atom. The site weights of the bands are consistent with it: the two highest valence bands lie on the Cr atoms of one layer (and on S and Br \(p\) states) and have almost no weight on the Cr atoms of the other layer, and the weight is shared equally by the two Cr atoms of the layer at nearly all \textit{k}-points, as for even and odd combinations (Supplementary Fig.~S2).} This view is the band-insulator analogue of familiar bright/dark symmetry-adapted excitons and Davydov-like splittings in molecular aggregates \cite{Davydov1971Excitons,Kasha1965Aggregates}.

\rev{This parity rule acts on the transitions at every \(k\), not only at a single high-symmetry \textit{k}-point.}

\section*{Discussion}

The resulting picture refines, rather than rejects, the ligand-field description. In a local ligand-field language, the low-energy excitons are orbital-orbital transitions within a Cr-centered \(d\)-manifold with ligand hybridization. This picture is valuable because it identifies the local orbital origin of the excitons and explains why purely onsite \dd\ transitions are dipole suppressed. However, the optical matrix element in a crystal acts on Bloch states. Each \(|vk\rangle\) and \(|ck\rangle\) already carries local orbital, ligand-hybridization, sublattice, and bonding/antibonding character.

Thus the correct physical object is not merely a local \(d{\rightarrow}d\) excitation. It is a symmetry-adapted band transition, a coherent superposition \(\sum_{kvc}A^S_{kvc}\,|vk\rangle\!\rightarrow\!|ck\rangle\) of Bloch-state transitions. The local ligand-field component tells us what orbitals participate; the band-symmetry component tells us whether the transition amplitudes add or cancel. \rev{Transitions that preserve the parity under the Cr exchange (the even and odd combinations of the two Cr atoms of a layer, that is, bonding and antibonding) are dipole-allowed along \(b\), and transitions that change it are not. Even excitons are therefore bright and odd excitons dark, even when both have similar local orbital content.}

This reconciles the competing descriptions of \CrSBr. The \(1.33\ev\) exciton is ligand-field-like and relatively localized, but its brightness is a band-coherent property. The \(1.84\ev\) exciton is more intersite and closer to a Wannier-Mott limit, but its brightness is still determined by its symmetry-adapted projection onto the optical vector. \CrSBr\ is therefore neither a simple molecular ligand-field system nor a conventional weakly bound Wannier exciton system. It is a magnetic band insulator in which ligand-field-like orbital excitations acquire optical strength through band hybridization, intersite coherence, and symmetry-adapted transition-dipole addition. This view sits naturally within the broader picture of 2D vdW magnets, where coupling between magnetic order, lattice, and optical excitations has emerged as a recurring organizing theme \cite{Park2026RMPReview,Adak2026Review}.

To recapitulate, the most important dark partners -- the \(\sim\!1.38\,\mathrm{eV}\) neighbour of \(X_A\) and the \(\sim\!1.82\,\mathrm{eV}\) neighbour of \(X_B\) -- sample nearly identical \textit{k}-space regions as their bright partners. \rev{They are dark because they are odd under the Cr exchange and the \(b\)-axis optical transition-density vector is nearly orthogonal to the odd sector.} The source of brightness in \CrSBr\ is therefore neither purely Frenkel-like onsite physics nor purely Wannier-Mott spatial extent. \rev{It is a parity selection rule of the structure, applied to coherent sums of band transitions within a ligand-field-derived magnetic band structure.} The same logic explains why these dark companions remain absent in equilibrium reflectivity and photoluminescence but appear in probes that bypass the optical matrix element or transiently relax the selection rule: the \(\sim\!1.5\,\mathrm{eV}\) dark exciton resolved by resonant inelastic x-ray scattering \cite{Sears2025RIXSDarkExciton} and the coherent magnon- and phonon-driven \(1.46\,\mathrm{eV}\) resonance in time-resolved reflectivity \cite{Borka2026OpticalInterface} \rev{lie in the energy window of the dark vertical states found here (1.38 to 1.53 eV), which includes a dark state at \(1.50\ev\). The bright/dark partitioning is a property of the excitonic manifold that experiments that bypass the optical matrix element can reach.}

\rev{The brightness hierarchy in \CrSBr\ is not controlled by onsite \dd\ weight alone. Bright and dark excitons are the even and odd members, under the exchange of the two Cr atoms of a layer, of coherent superpositions of ligand-field-like and charge-transfer-like band transitions drawn from similar bands and a similar region of \textit{k}-space. The parity fixes whether the \(b\)-axis dipole is allowed, and the coherent addition of the allowed transitions sets how bright an allowed state is.}

\section*{Methods}

\paragraph*{Electronic structure.} The excitonic spectrum of bulk \CrSBr\ in the antiferromagnetic ground state was computed within self-consistent many-body perturbation theory. The starting-point-independent quasiparticle self-consistent \(GW\) framework \cite{vanSchilfgaarde2006QSGW,Kotani2007QSGW} was first iterated to self-consistency in the static, Hermitised self-energy \(\Sigma_0\); ladder (electron-hole) diagrams were then added to the polarisability used to build the screened interaction \(W\), and the full \QSGW\ cycle \cite{Cunningham2023QSGWhat} was iterated to self-consistency in both \(\Sigma\) and the charge density. Inclusion of the electron-hole vertex in \(W\) is essential for both the quasiparticle gap and the optical response of magnetic insulators \cite{Acharya2021CrX3Bands,Acharya2022CrX3}. \rev{The antiferromagnetic bulk cell contains 12 atoms (4 Cr, 4 S, 4 Br) with \(a=3.504\,\)\AA, \(b=4.738\,\)\AA\ and \(c=15.81\,\)\AA.} Single-particle quantities were constructed on a \(10\times 7\times 2\) \textbf{k}-mesh and converged to an r.m.s.\ change of \(\Sigma_0\) below \(10^{-5}\,\)Ry.

\paragraph*{Excitons.} The two-particle Hamiltonian entering the Bethe-Salpeter equation \cite{SalpeterBethe1951,Strinati1988BSE,Onida2002BSE} was built from \(26\) valence and \(9\) conduction bands per \textbf{k}-point on the same \(10\times 7\times 2\) mesh (symmetry-unfolded), giving a per-spin rank of \(32\,760\) transition components; eigenvalues, eigenvectors \(A^{S}_{kvc}\), and oscillator strengths in all three polarisation channels were extracted from the converged \BSE\ solutions. \rev{The window was chosen to contain all Cr-\(d\) and S- and Br-\(p\) derived states relevant to the low-energy excitons. With four Cr\(^{3+}\) (\(t_{2g}^3\)) ions per cell, the occupied Cr-\(d\) states provide 12 valence bands and the empty Cr-\(d\) states 8 conduction bands, and the S and Br \(p\) states are included on the same footing. We explored the number of bands over the whole space of these states, and beyond the set of 26 valence and 9 conduction bands the exciton energies stopped changing or moved by about \(1\) meV. We therefore used this window for all calculations, as in our earlier work \cite{Datta2025MagnonMediated}.} \rev{The \BSE\ is solved in the Tamm-Dancoff approximation with the statically screened interaction of the \QSGW\ calculation, and without spin-orbit coupling; calculations that include it do not change the essential results for \CrSBr, in contrast to \(\mathrm{CrI}_3\) \cite{Acharya2022CrX3}. The calculation is for the full two-layer antiferromagnetic cell with collinear order, and the two spin sectors are solved separately; their \BSE\ eigenvalues come out degenerate, as the symmetry of the cell (a spin flip together with the exchange of the two layers) requires. The quasiparticle gap is about \(2.1\ev\).}

\paragraph*{Interference figure of merit.} The per-channel transition-density vector \(\rho^{\alpha}_{kvc}\) of Eq.~\eqref{eq:rho} was reconstructed from the same single-particle eigenvalues and optical matrix elements. The coherent reconstruction \(|\sum A^S\rho|^2\) reproduces the \BSE\ oscillator strengths to numerical precision across the full spectrum, in all three polarisations and both spin sectors; the ratio of this coherent sum to the incoherent sum \(\sum|A^S\rho|^2\) gives Eq.~\eqref{eq:cohinc}. \rev{The largest deviation of the reconstruction from the oscillator strengths reported by the \BSE\ solver is \(3\times10^{-13}\) of the strongest line. States with an oscillator strength below \(10^{-8}\) of the maximum are at the numerical resolution of the sums, so values of \(\mathcal I\) below about \(10^{-8}\) mean only that the state is dark. The oscillator strength of the symmetry-forbidden \(1.38\ev\) partner of \(X_A\), relative to \(X_A\), is \(2.7\times10^{-8}\) on the \(10\times7\times2\) mesh and stays at about \(2\times10^{-8}\) on \(10\times8\times2\) and \(11\times8\times2\) meshes.}

\paragraph*{Auxiliary diagnostics.} The \textit{k}-space participation ratio reported in Table~\ref{tab:states} is \(k\mathrm{PR}(S)=1/\sum_k P_S(k)^2\) with \(P_S(k)=\sum_{vc}|A^{S}_{kvc}|^2\); the parity weights follow Eq.~\eqref{eq:weven}; the k-overlap quoted between near-degenerate excitons is the Bhattacharyya overlap of \(P_S(k)\) and \(P_{S'}(k)\)\rev{, \(\mathrm{BC}(S,S')=\sum_k\sqrt{P_S(k)P_{S'}(k)}\) \cite{Kailath1967Bhattacharyya}, which is 1 for identical distributions}

\rev{\paragraph*{Symmetry analysis.} The symmetry operations are the orthorhombic point operations with a fractional translation that map the 12-atom cell, including the Cr spin labels, onto itself. Each was represented on the Bloch basis of the calculation from the atom permutation, the parity of each real harmonic and the Bloch phases of the atom positions. The representation is consistent with the structure: the basis overlap matrix is invariant under all eight operations, the relative parity of every transition is \(\pm1\) to numerical precision at all \(k\), and the exciton characters are \(\pm1\) to within \(5\times10^{-3}\) for the states of Table~\ref{tab:states}. Bands closer than \(0.5\) meV are treated as degenerate, and an exciton is counted as classified if \(|\chi|\) is within \(0.05\) of 1 for all operations.}

\section*{Data availability}

\rev{The inputs and the converged outputs of the \QSGW\ and \BSE\ calculations of this paper are publicly available on Zenodo, \href{https://doi.org/10.5281/zenodo.21880186}{doi:10.5281/zenodo.21880186}. The record contains the Questaal control, basis, site and SCF-restart files, the batch scripts of the \QSGW+\BSE\ workflow, the converged \QSGW\ self-energy, and the \BSE\ output files with the band-, orbital- and real-space-resolved weights of the excitons discussed here.}

\section*{Author contributions}

S.A.\ conceived the work, developed the methodology, performed and analysed the calculations, and wrote the manuscript. J.M., D.P., M.v.S., J.C.J. and J.L.B.  contributed to the interpretation of the results and to editing the manuscript.

\section*{Competing interests}

The authors declare no competing financial or non-financial interests.

\section*{Acknowledgements}
This work was authored by the National Laboratory of the Rockies for the U.S. Department of Energy (DOE) under Contract No.\ DE-AC36-08GO28308. S.A., J.M., J.C.J., and J.L.B. acknowledge funding from the Laboratory Discretionary Research and Development (LDRD) program of the National Laboratory of the Rockies.  For M.v.S. and D.P., funding was provided by the Computational Chemical Sciences program within the Office of Basic Energy Sciences, U.S.\ Department of Energy. S.A.\ acknowledges the use of computational resources sponsored by the Department of Energy's Office of Energy Efficiency and Renewable Energy and located at the National Laboratory of the Rockies. S.A. acknowledges the use of the National Energy Research Scientific Computing Center, under Contract No. DE-AC02-05CH11231 using NERSC award BES-ERCAP0021783 for a portion of the work. The views expressed in the article do not necessarily represent the views of the DOE or the U.S.\ Government. The U.S.\ Government retains and the publisher, by accepting the article for publication, acknowledges that the U.S.\ Government retains a nonexclusive, paid-up, irrevocable, worldwide license to publish or reproduce the published form of this work, or allow others to do so, for U.S.\ Government purposes.

\section*{Funding}
This work was authored by the National Laboratory of the Rockies for the U.S. Department of Energy (DOE) under Contract No.\ DE-AC36-08GO28308. S.A., J.M., J.C.J., and J.L.B. acknowledge funding from the Laboratory Discretionary Research and Development (LDRD) program of the National Laboratory of the Rockies. For M.v.S. and D.P., funding was provided by the Computational Chemical Sciences program within the Office of Basic Energy Sciences, U.S.\ Department of Energy.

\clearpage
\onecolumngrid
\begin{center}\large\textbf{Supplementary Information}\end{center}
\bigskip
\setcounter{figure}{0}\setcounter{table}{0}
\renewcommand{\thefigure}{S\arabic{figure}}\renewcommand{\thetable}{S\arabic{table}}

\noindent This Supplementary Information accompanies the main text and documents the band, orbital and real-space content of the excitons discussed there. Fig.~S1 gives the band-resolved weight of all 11 excitons of Table~I. Fig.~S2 shows the orbital character of the bands and the site character of the excitons, and Table~S2 the Cr \(d\)-orbital and ligand character of the hole and of the electron. Fig.~S3 compares the \BSE\ dielectric function with the sum over the lowest excitons. Fig.~S4 shows the real-space electron density for a hole fixed on a Cr atom. Fig.~S5 shows the dipole matrix elements of the four leading transitions along two high-symmetry lines, and Table~S1 the weight of the excitons on the four-band block. Notation, units and symmetry labels are those of the main text.

\begin{figure}[!htbp]
 \centering
 \includegraphics[width=\textwidth]{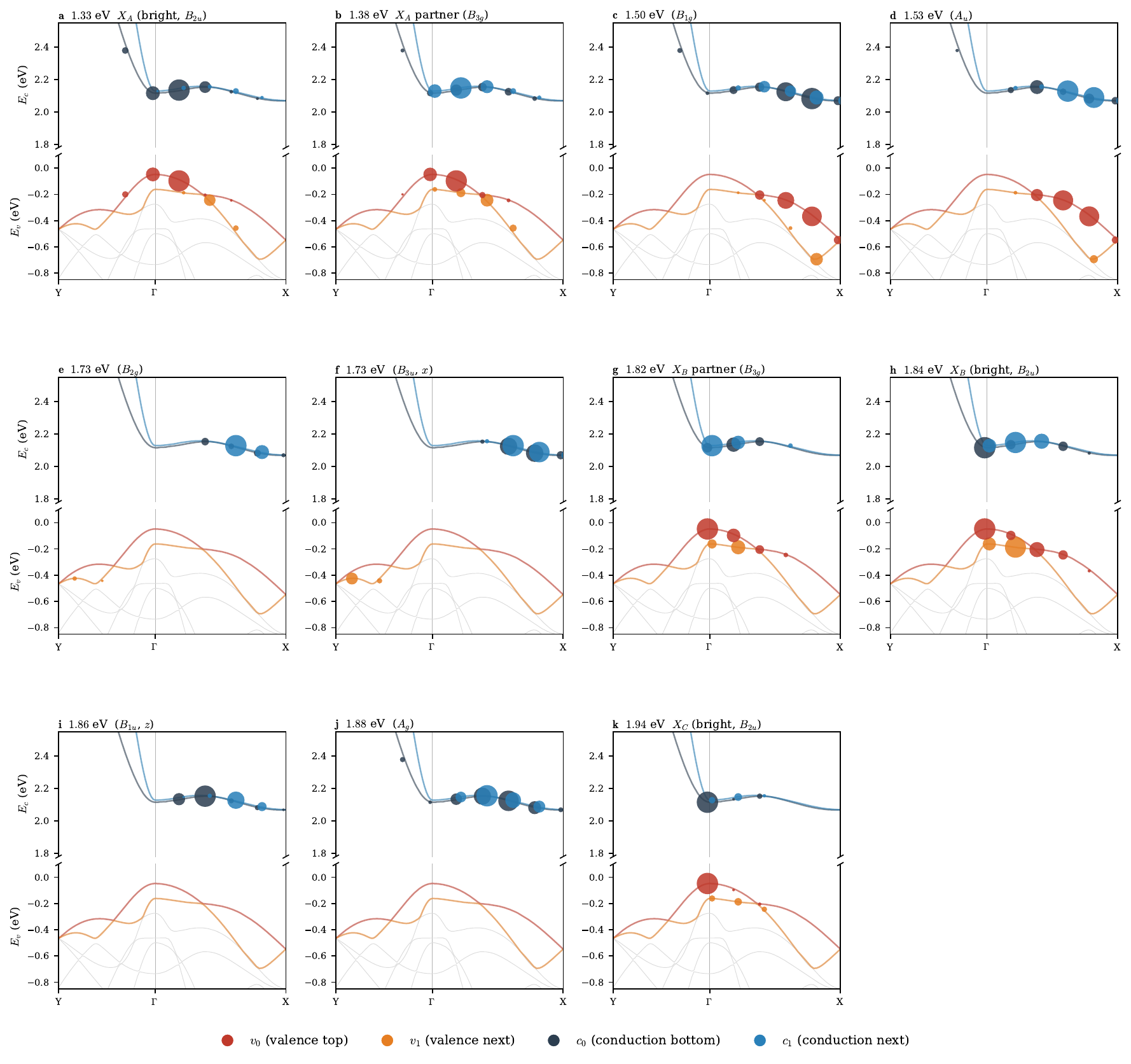}
 \caption{\rev{\textbf{Band-resolved \BSE\ weight of all 11 excitons of Table~I.} Panels a) to k) show the excitons at a) \(1.33\ev\) (bright \(X_A\), \(B_{2u}\)), b) \(1.38\ev\) (\(X_A\) partner, \(B_{3g}\)), c) \(1.50\ev\) (\(B_{1g}\)), d) \(1.53\ev\) (\(A_u\)), e) \(1.725\ev\) (\(B_{2g}\)), f) \(1.728\ev\) (\(B_{3u}\), \(x\)-polarised), g) \(1.82\ev\) (\(X_B\) partner, \(B_{3g}\)), h) \(1.84\ev\) (bright \(X_B\), \(B_{2u}\)), i) \(1.86\ev\) (\(B_{1u}\), \(z\)-polarised), j) \(1.88\ev\) (\(A_g\)) and k) \(1.94\ev\) (bright \(X_C\), \(B_{2u}\)); the symbols in brackets are the irreducible representations of \(D_{2h}\). Each panel shows the weight along \(\Gamma\)-X and Y-\(\Gamma\) (mesh points with \(k_z=0\)) on a broken energy axis: the upper part shows the two lowest conduction bands \(c_0\) (dark blue) and \(c_1\) (light blue) with energy \(E_c\), the lower part the two highest valence bands \(v_0\) (red) and \(v_1\) (orange) with energy \(E_v\), ordered by energy at each \(k\); grey lines are other bands. Marker area scales with the square of the weight relative to the largest weight of the four-band block, as in Fig.~3 of the main text.}}
 \label{fig:sm1}
\end{figure}

\begin{figure}[!htbp]
 \centering
 \includegraphics[width=0.95\textwidth]{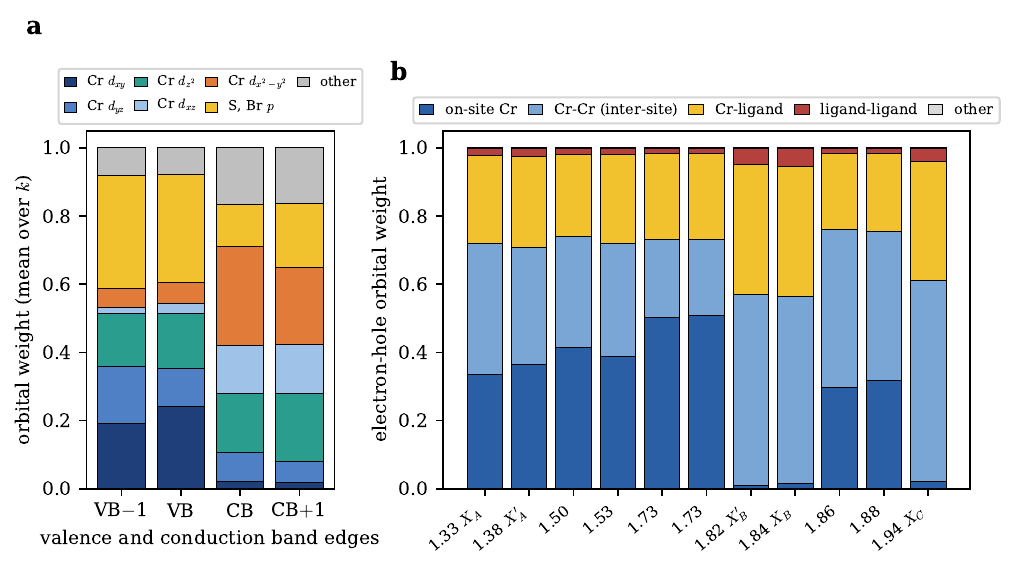}
 \caption{\rev{\textbf{Orbital character of the bands and site character of the excitons.} a) Mean weight over the mesh of the two highest valence bands (VB\(-1\), VB) and the two lowest conduction bands (CB, CB\(+1\)) on the individual Cr \(d\) orbitals (crystal axes, \(x\parallel a\), \(y\parallel b\), \(z\parallel c\)), on S and Br \(p\) orbitals and on the rest (other). The weights are Mulliken weights computed from the band eigenvectors and the overlap matrix, summed over the muffin-tin basis functions of the atoms. Colours: dark blue \(d_{xy}\), blue \(d_{yz}\), teal \(d_{z^2}\), light blue \(d_{xz}\), orange \(d_{x^2-y^2}\), yellow S and Br \(p\), grey other. b) Weight of each of the 11 excitons of Table~I by site class, as reported by the \BSE\ code: electron and hole on the same Cr atom (on-site, dark blue), Cr-Cr (all other pairs, including the same atom in a different cell, light blue), Cr-ligand (yellow), ligand-ligand (red) and the remainder (other, grey), which is weight on empty-sphere channels. Excitons are labelled by their energy in eV.}}
 \label{fig:sm2}
\end{figure}

\begin{figure}[!htbp]
 \centering
 \includegraphics[width=0.55\textwidth]{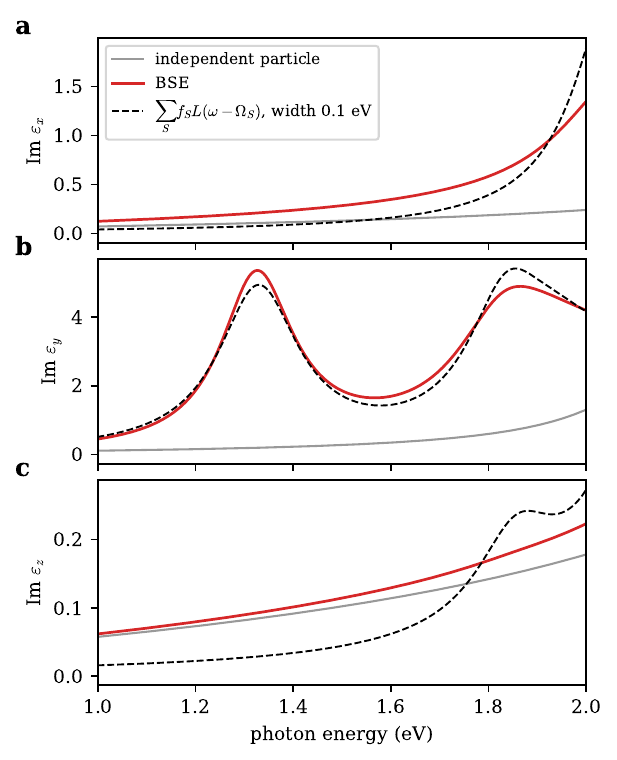}
 \caption{\rev{\textbf{Imaginary part of the dielectric function along the three crystal axes.} Panels a), b) and c) show Im\(\,\varepsilon_x\), Im\(\,\varepsilon_y\) and Im\(\,\varepsilon_z\) (\(x\parallel a\), \(y\parallel b\), \(z\parallel c\)) against photon energy. In each panel the red line is the \BSE\ calculation and the grey line the independent-particle spectrum. The black dashed line places a Lorentzian of width \(0.1\ev\) at each of the 300 lowest excitation energies \(\Omega_S\) and weights it by the oscillator strength \(f_S\) of that line, with one scale factor per panel. Along \(b\) (panel b) this sum follows the \BSE\ line shape from \(1.0\) to \(2.0\ev\). Along \(a\) and \(c\) (panels a and c) the same window holds only a weak background, and those 300 lines do not account for it.}}
 \label{fig:sm3}
\end{figure}

\begin{figure}[!htbp]
 \centering
 \includegraphics[width=0.95\textwidth]{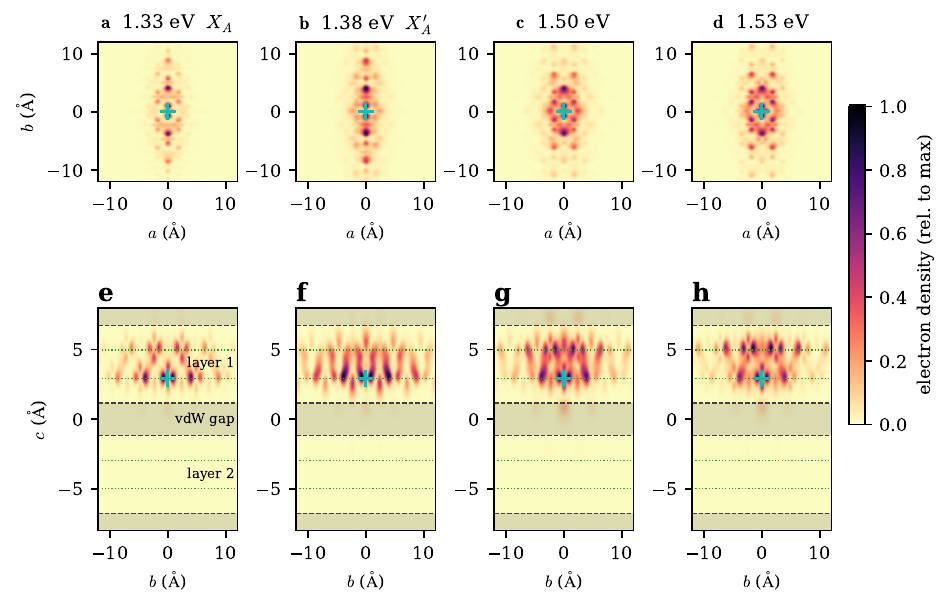}
 \caption{\rev{\textbf{Real-space electron density for a hole fixed on a Cr atom.} The hole is fixed on a Cr atom of layer 1 (cyan cross). Panels a) to d) show the electron density integrated over layer 1 and viewed along \(c\), for a) \(X_A\) (\(1.33\ev\)), b) its partner \(X_A'\) (\(1.38\ev\)), c) the state at \(1.50\ev\) and d) the state at \(1.53\ev\). Panels e) to h) show the same four states integrated over \(a\) and viewed along \(a\): e) \(1.33\ev\), f) \(1.38\ev\), g) \(1.50\ev\), h) \(1.53\ev\). In e) to h), dashed lines mark the outer Br planes of each layer, dotted green lines the two Cr planes, and the shaded bands the van der Waals gaps between layer 1 and layer 2. The colour scale gives the electron density relative to its maximum, from 0 (light yellow) to 1 (black). Between \(97\%\) and \(99\%\) of the electron density lies in layer 1.}}
 \label{fig:sm4}
\end{figure}

\begin{figure}[!htbp]
 \centering
 \includegraphics[width=0.95\textwidth]{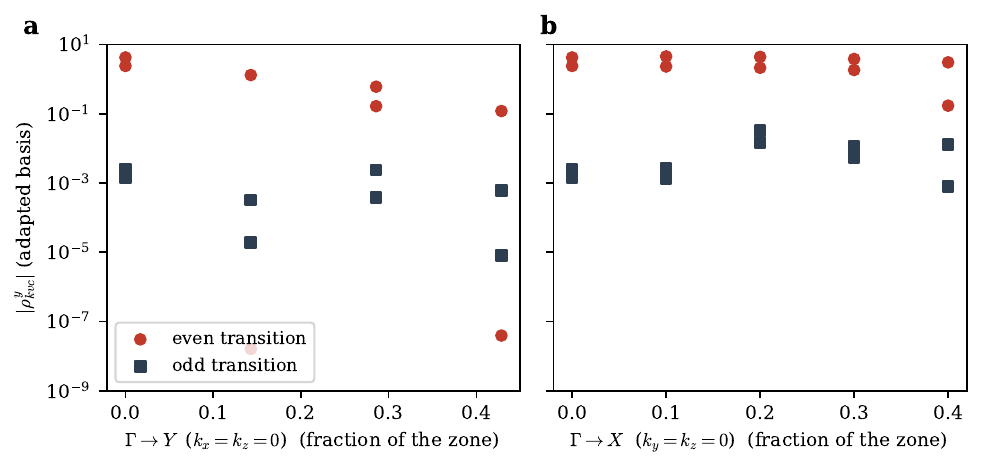}
 \caption{\rev{\textbf{Dipole matrix elements of the four leading transitions along two high-symmetry lines.} Magnitude of the \(b\)-axis transition vector \(|\rho^{y}_{kvc}|\) (logarithmic scale) of the four transitions between the two highest valence and the two lowest conduction bands, in the symmetry-adapted basis, at the mesh points of a) the \(\Gamma\)-Y line (\(k_x=k_z=0\)) and b) the \(\Gamma\)-X line (\(k_y=k_z=0\)); the horizontal axis is the distance from \(\Gamma\) as a fraction of the zone. Red circles are transitions that are even and dark blue squares transitions that are odd under \(m_z\). The even transitions are large, the odd transitions are suppressed by two to four orders of magnitude, with one accidental even-sector minimum at one point of \(\Gamma\)-Y.}}
 \label{fig:sm5}
\end{figure}

\begin{table}[!htbp]
 \caption{\rev{Weight of the excitons of Table~I on the four-band block (the two highest valence and two lowest conduction bands), on the two lowest valence bands of the active window and on the two highest conduction bands of the window. Weights sum to one over the whole window.}}
 \label{tab:s1}
 \begin{ruledtabular}
 \begin{tabular}{cccc}
 \(E\) (eV) & four-band block & lowest valence bands & highest conduction bands \\
 \hline
 \input{table_S1_rows.tex}
 \end{tabular}
 \end{ruledtabular}
\end{table}

\begin{table}[!htbp]
 \caption{\rev{Cr \(d\)-orbital and ligand character of the hole and of the electron of the excitons of Table~I, in the crystal axes (\(x\parallel a\), \(y\parallel b\), \(z\parallel c\)). Hole: weight on Cr \(d_{yz}\), Cr \(d_{xy}\) and S, Br \(p\) orbitals; electron: weight on Cr \(d_{x^2-y^2}\), Cr \(d_{z^2}\) and S, Br \(p\) orbitals. The weights are the orbital weights of the valence (hole) and conduction (electron) bands, averaged with the squared amplitudes of the exciton; they neglect phase coherence between bands and are an estimate. The remainder of each weight lies on the other Cr \(d\) orbitals, other channels and other atoms.}}
 \label{tab:s2}
 \begin{ruledtabular}
 \begin{tabular}{ccccccc}
  & \multicolumn{3}{c}{hole} & \multicolumn{3}{c}{electron} \\
 \(E\) (eV) & \(d_{yz}\) & \(d_{xy}\) & S, Br \(p\) & \(d_{x^2-y^2}\) & \(d_{z^2}\) & S, Br \(p\) \\
 \hline
 \input{table_S2_rows.tex}
 \end{tabular}
 \end{ruledtabular}
\end{table}

\end{document}

%% file: table_S1_rows.tex
1.33 & 0.652 & $1.8\times10^{-4}$ & $0.013$ \\
1.38 & 0.533 & $2.4\times10^{-4}$ & $0.017$ \\
1.50 & 0.448 & $3.0\times10^{-4}$ & $0.019$ \\
1.53 & 0.511 & $2.8\times10^{-4}$ & $0.019$ \\
1.73 & 0.212 & $1.0\times10^{-3}$ & $7.1\times10^{-3}$ \\
1.73 & 0.203 & $1.0\times10^{-3}$ & $6.9\times10^{-3}$ \\
1.82 & 0.954 & $2.2\times10^{-5}$ & $9.7\times10^{-5}$ \\
1.84 & 0.961 & $2.6\times10^{-5}$ & $1.0\times10^{-3}$ \\
1.86 & 0.250 & $2.6\times10^{-4}$ & $5.9\times10^{-3}$ \\
1.88 & 0.145 & $2.3\times10^{-4}$ & $7.4\times10^{-3}$ \\
1.94 & 0.945 & $1.9\times10^{-5}$ & $1.1\times10^{-3}$ \\

%% file: table_S2_rows.tex
1.33 & 0.44 & 0.09 & 0.34 & 0.47 & 0.19 & 0.12 \\
1.38 & 0.44 & 0.10 & 0.34 & 0.45 & 0.15 & 0.16 \\
1.50 & 0.15 & 0.43 & 0.29 & 0.40 & 0.18 & 0.16 \\
1.53 & 0.14 & 0.38 & 0.33 & 0.40 & 0.19 & 0.15 \\
1.73 & 0.08 & 0.09 & 0.27 & 0.07 & 0.03 & 0.24 \\
1.73 & 0.09 & 0.09 & 0.26 & 0.06 & 0.03 & 0.25 \\
1.82 & 0.36 & 0.05 & 0.49 & 0.54 & 0.16 & 0.12 \\
1.84 & 0.31 & 0.08 & 0.51 & 0.54 & 0.15 & 0.12 \\
1.86 & 0.02 & 0.03 & 0.31 & 0.47 & 0.22 & 0.12 \\
1.88 & 0.05 & 0.02 & 0.29 & 0.47 & 0.20 & 0.13 \\
1.94 & 0.45 & 0.03 & 0.42 & 0.55 & 0.19 & 0.09 \\